\documentclass[onecolumn,authoryear]{els-mrw} 

\usepackage{amsmath,amssymb,amsfonts,amsthm,makeidx,graphicx}
\usepackage{txfonts}
\usepackage{helvet}
\usepackage{ulem}
\usepackage{hyperref}
\usepackage{minitoc}
\newcommand{\grl}{Geophysical Research Letters}
\newcommand{\aap}{Astronomy \& Astrophysics}
\newcommand{\apj}{The Astrophysical Journal}

\newcommand{\nat}{Nature}
\newcommand{\apjl}{The Astrophysical Journal Letters}
\newcommand{\jgr}{Journal of Geophysical Research}

\newcommand{\icarus}{Icarus}
\newcommand{\mnras}{Monthly Notices of the Royal Astronomical Society}
\newcommand{\planss}{Planetary and Space Science}
\newcommand{\psj}{The Planetary Science Journal}
\newcommand{\ssr}{Space Science Reviews}
\newcommand{\aj}{The Astronomical Journal}

\begin{document}
\dominitoc
\fontsize{12}{15}\selectfont
\chapter{Ice Giants}\label{chap1}

\author[1]{Ravit Helled}%

\address[1]{\orgname{Department of Astrophysics}, \orgdiv{University of Zurich}, \orgaddress{Winterthurerstr. 190, 8057 Zurich, Switzerland}}


\articletag{Chapter Article tagline: update of previous edition,, reprint..}

\maketitle

\section{Summary}

{\large Uranus and Neptune, the so-called "ice giants"  in the solar system, represent a fascinating class of planets. They are the farthest planets in the solar system with intermediate masses/sizes, complex non-polar magnetic fields, strong atmospheric winds, and not well-understood internal structures. 
Studying the interiors of Uranus and Neptune  is vital for advancing our understanding of planetary formation and evolution as well as for the characterization of planets around other stars. 
In this review, we summarize our current knowledge of the interior and formation of Uranus and Neptune. The internal structures and compositions of Uranus and Neptune are derived from numerical models that fit the available measurements. 
Both planets are expected to be composed of rocks and ices and have H-He atmospheres of the order of 10\% of their total masses.  The rock-to-water ratios in Uranus and Neptune, however,  are very uncertain. It is also unclear how the different materials are distributed within the interiors and whether distinct layers exist. 
While often Uranus and Neptune are viewed as being "twin planets" it is in fact unclear how different the two planets are from each other, and whether they are indeed "icy" (water-dominated) 
planets. After summarizing the current-knowledge of the interiors of Uranus and Neptune, we briefly discuss their magnetic fields and atmosphere dynamics.  
We next introduce the challenges in constraining the formation paths of Uranus and Neptune: it is still unclear whether the planets formed at their current locations, and what the dominating processes that led to their formation (accretion rates, size of solids, etc.) were. We also mention the possible role of giant impacts shortly after their formation. 
Finally, we suggest that advanced modeling, future observations from space and the ground, lab experiments, and links with exoplanetary science can improve our understanding of Uranus and Neptune as a class of astronomical objects which seems to be very common in our galaxy. }

\subsection*{Keywords:} {\large Uranus, Neptune, ice giants, gaseous planets, interior models, planetary structure}


\minitoc

\newpage
\section{The internal structures of Uranus and Neptune}

Uranus and Neptune, also known as the "ice giants", are the outermost  planets in the Solar System, with semi-major axes of 19.19 AU and 30.07 AU\footnote{The Astronomical Unit (AU) is a unit of length corresponding  to the average distance between the Sun and Earth.}, respectively. These planets are still not well understood and currently, there are still many fundamental mysteries surrounding them, including their origin, evolution, overall composition and internal structure, rotation rates, magnetic fields, and atmosphere dynamics.  
The masses of Uranus and Neptune are 14.5 and 17 Earth mass (M$_{\oplus}$), respectively, and their sizes are about 3.98 and 3.86 times Earth's radius. The average densities of the planets (inferred from the mass and radius measurements) are 1.27 g cm$^{-3}$ and 1.64 g cm$^{-3}$ for Uranus and Neptune, respectively. These densities indicate that a large fraction of the planetary interior consist of elements heavier than hydrogen-helium (H-He). The heavier elements\footnote{All the elements heavier than helium are referred to as "heavy elements".} are typically divided into volatile materials, or "ices" (water, ammonia, and methane), or refractory materials ("rocks").   Formation and interior models predict that the mass fraction of H-He in the planets is of the order of only $\sim 10-15\%$. 
\par

Our understanding of the ice giants and their systems (rings, moons) is incomplete mainly due to the limited available measurements. Both Uranus and Neptune were only visited by the spacecraft Voyager 2 in the mid-1980s as it flew by  the planets. While these flybys provided key measurements they were limited due to the short time the spacecraft spent near the planets. From the ground it is very challenging to observe the planets due to their large distances from Earth. In addition, as more exoplanets have been detected and characterized, it became clear that planets with sizes/masses similar to those of the ice giants are extremely common in the galaxy. 
This adds motivation to understand Uranus and Neptune that belong to a unique planetary type, which is clearly  different from terrestrial planets and gas giant planets.   
\par  

Efforts to determine the internal structure of Uranus and Neptune go decades back \citep[e.g.,][]{1965JGR....70..199R,Podolak1995,Marley1995}. 
The bulk composition and internal structure of Uranus and Neptune must be inferred from interior structure models. As discussed below, there are two type of interior models that are used to investigate the planetary interior: "physical models" and "empirical  models". 

\subsection{Standard Models}
Physical interior models  solve the standard structure equations assuming a certain composition with an equation of state (EOS) of the assumed materials and aim to fit the planetary observed properties. These typically correspond to the planet's mass, radius, gravitational
field, 1-bar temperature, atmospheric
composition (when available), and rotation rates. 
The key physical properties of Uranus and Neptune are listed in Table \ref{tab:properties_planets}. 
In particular, the measured gravity field is used to constrain the density distribution inside the planet. 
Measurements of the gravitational field can be translated into gravitational moments, $J_n$, and interior models are designed to reproduce the measured moments (see \cite{HelledHoward2024} for further details on the construction of structure models). 
For Uranus and Neptune  only $J_2$ and $J_4$ have been measured with relatively large uncertainties on their values which leads to large uncertainties in their inferred structures and bulk compositions.

\begin{figure}[h]
    \centering
\includegraphics[width=0.65\paperwidth]{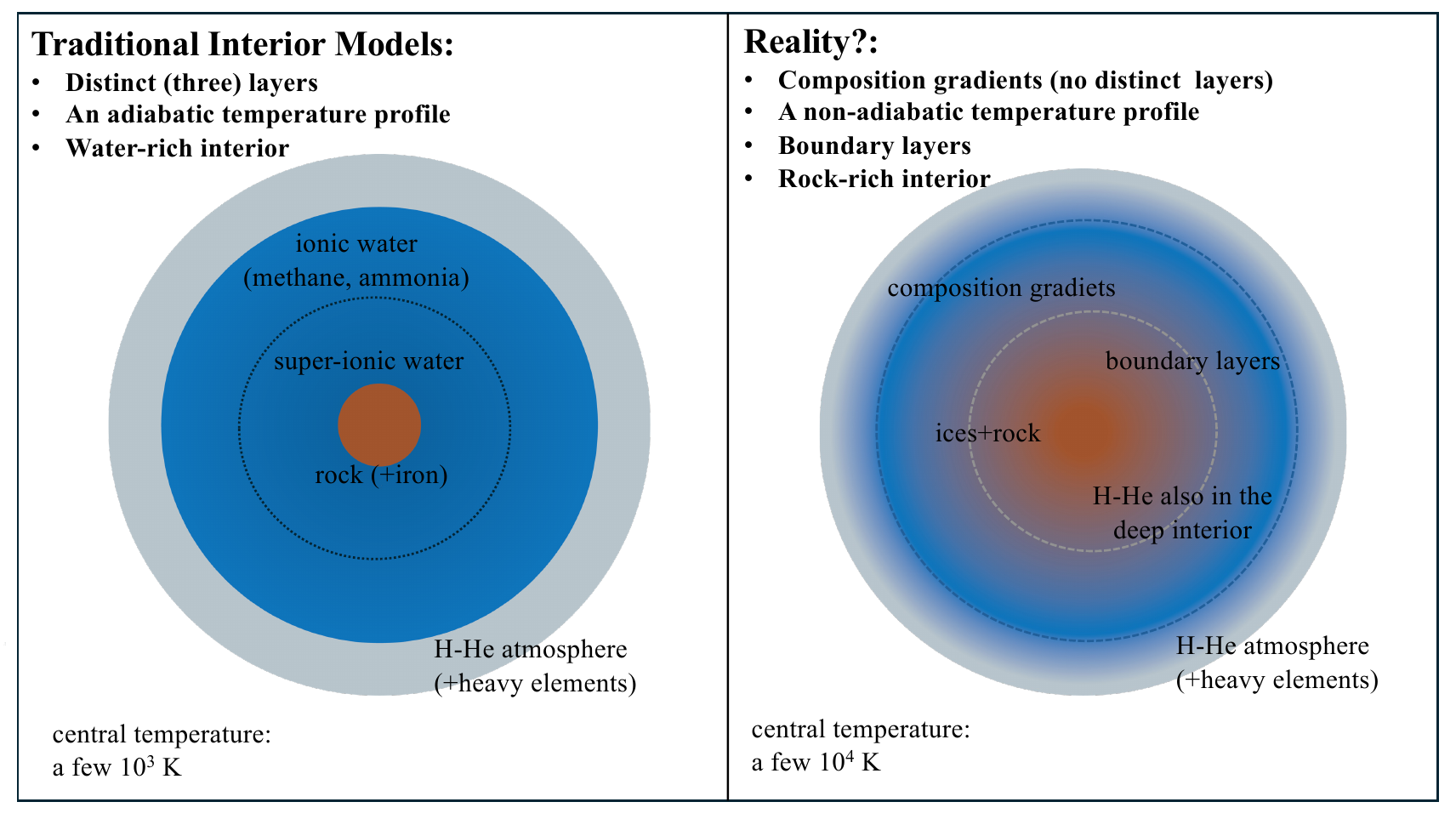}
    \caption{Sketches of the internal structures of the ice giants.  {\bf Left}: Traditional three-layer models that include a rocky core, water-rich inner envelope, and an atmosphere. Such models typically assume an adiabatic interior and result in central temperatures of the order of 10$^3$ K.  {\bf Right:} A more complex internal structure that includes composition gradients, boundary layers, and  mixtures of water and rock, as well as the existence of H-He in the deep interior. Such a scenario implies that the planets would also have non-convective regions and higher  central temperatures. Note that in both cases ionic water is expected to transition to super-ionic water \citep{redmer}.}
    \label{fig:rho}
\end{figure}

Standard structure models of Uranus and Neptune assume that the planetary interior is divided into three-layers including a rocky core, water envelope, and a H-He atmosphere enriched with heavy-elements \cite[e.g.,][]{nettel13}. In such models, also known as "physical models", the  species existing  in the planet must be assumed a priori, and typically it is assumed that the planets are adiabatic, i.e., that the temperature profile follows adiabatic temperature gradient\footnote{The adiabatic gradient
$\nabla_{ad}=\frac{d ln{T}}{d ln{P}}\arrowvert_S$, where $T$ and $P$ are the temperature and pressure, respectively, and $S$ is the
entropy.}, and that each layer is homogeneous in composition. As a result, these physical/traditional models are also referred to as "adiabatic models". 
It should be noted that these models are simple in the sense that they assume a differentiated interior and rely on the modeler's guess on the expected materials in the planetary interior. In a way, these traditional models are guided by our knowledge of Earth (which is differentiated) and the structures that are expected from giant planet formation models.
\par 

Three-layer models of Uranus suggest that the planet consists of  up to  $\sim$ 2 $M_{\oplus}$ of H-He and that the outer envelope metallicity (heavy element mass fraction, $Z$) is $\leq10-20\%$ (depending on the assumed rotation rate). The inner envelope is expected to have  90\%  of heavy elements. 
Such models predict that Neptune has an outer envelope metallicity of $\sim$ 55-65\%, depending on the assumed rotation period although solution with envelopes of solar metallicity are also possible. The rocky/metal cores are found to have masses ranging from nearly zero to about 5 $M_{\oplus}$ (see \cite{nettel13} for details).

\begin{table}[h!]
{\small 
\def\arraystretch{1.2}
\centering 
\begin{tabular}{lll}
\hline
\hline
{\bf Parameter} &  {\bf Uranus} & {\bf Neptune}\\
\hline
Semi-major axis$^a$ (AU) &   19.19126393  & 30.06896348   \\
Mass ($10^{24}$ kg)   & 86.8099 ±0.0040  & 102.4092 ±0.0048 \\
Mean Radius (km)&  25362 ±7 & 24622 ±19  \\
Equatorial Radius (km) & 25559 ±4  &24764 ±15 \\
Mean Density (g cm$^{-3}$)$^b$  & 1.270 $\pm$ 0.001 & 1.638 $\pm$ 0.004 \\
R$_{\rm ref}$$^c$ (km)  & 25,559 & 24,764 \\
J$_2$ ($\times$10$^{6}$)   &3510.68 $\pm$ 0.70& 3408.43 $\pm$ 4.50 \\
J$_4$ ($\times$10$^{6}$)  & $-$34.17 $\pm$ 1.30 & $-$33.40 $\pm$ 2.90 \\
J$_6$ ($\times$10$^{6}$) &  46.12--59.90$^d$  & 45.26--68.63$^d$ \\
J$_8$ ($\times$10$^{6}$) & -8.4 to -17.8$^d$& -7.85 to  -20.16$^d$\\
Moment of inertia (I/MR$^2$) & 0.225 & 0.23-0.25$^d$ \\
Rotation period$^d$ (hr) & 17.24  & 16.11 \\
1-bar Temperature (K)  & 76 $\pm$ 2& 72 $\pm$ 2\\
Emitted power ($10^{16}~$J/s) & 0.560 $\pm$ 0.011& 0.534 $\pm$ 0.029\\
Absorbed power ($10^{16}~$J/s) & 0.526 $\pm$ 0.037& 0.204 $\pm$ 0.019\\
Bond Albedo $A$ & 0.300 & 0.290 \\
\hline
\hline
\end{tabular}
\caption{Basic physical properties of Uranus and Neptune (from \url{https://nssdc.gsfc.nasa.gov/planetary/} and \url{https://ssd.jpl.nasa.gov/planets/phys_par.html}). $^a$J2000. $^b$The bulk Density computed based on the volume of a sphere with the published mean radius. $^c$R$_{\rm ref}$ is the reference equatorial radius in respect to the measured gravitational harmonics. $^d$Calculated values from \citep{Neuenschwander2022}. 
$^e$Note that the rotation periods of the planets are not well determined, in particular in the case of Uranus and Neptune \citep{Helled2010shape}. $^f$The value from Voyager is displayed. 
}}
\label{tab:properties_planets}       
\end{table}
In reality, the parameter space of possible internal structures and chemical compositions is huge. It is indeed expected that the internal structures of Uranus and Neptune include composition gradients, non-convective regions and boundary layers. Composition gradients in the planetary deep interior are predicted by formation models and expected to be caused by the formation process where heavy elements (solids) are accreted along with the H-He gas \citep{Helled2017}. 
Boundary layers could also form due to phase separation and immicibilities that occur in the planetary interior \cite[e.g.,][]{Scheibe2021,2021PSJ.....2..222S}. Figure 1 shows traditional interior models of Uranus and Neptune (left) vs.~more realistic interiors (right). 
\par

Recently, more complex physical interior models of Uranus and Neptune have been presented. 
\cite{Bailey2021} investigated how demixing of water and hydrogen affects the internal structures of the planets. They found that Uranus' envelope metallicity is $\lesssim$  0.01, potentially suggestive of fully demixed hydrogen and water while for Neptune they found an  envelope that 10 times more water-rich. 
 \cite{Scheibe2021} presented a Uranus model with boundary layers which lead to  hotter interior than predicted by simple adiabatic models, with central temperatures beyond $10^4$ K. 
  Neptune could also have boundary layers, and it was shown that their thicknesses significantly affect the inferred internal structure. \citet{Vazan2020} suggested that composition gradients in Uranus' deep interior can explain  its observed luminosity. In such a scenario, which is supported by formation models (see below), the interior is hotter, and the inferred metalicity can be higher.  
Note that in some of these models the planetary composition and internal structure is determined from evolution models which are designed to fit the planetary luminosity at the planet's current age (see the review by N.~Nettelmann for further details). 

\subsection{Empirical Models}
 An alternative approach to "physical models" is to parameterize the planetary density distribution (using a mathematical/numerical function) and search for solutions that only match the measured mass, radius, and gravitational  field. This framework is known as “empirical models” or "agnostic models".  
 Given that the compositions of Uranus and Neptune are poorly constrained, the use of empirical  models allows us to investigate the planetary internal structures without relying on physical equations of state. The advantage of this modeling approach is that its inferred density profile could include solutions that are missed by conventional models (which typically have a simplified internal structure and an adiabatic interior).  
The disadvantage is that the inferred density profile may be  nonphysical. 
Some solutions, however, can be dismissed at a later stage if a compositional interpretation is performed. For empirical models, the composition can be determined by taking the inferred density profile combined with an assumed temperature profile, and then by using a physical EOS (as the pressure is determined by the hydrostatic equation). Then, the planetary composition and internal structure can be derived.  Although the modeler still relies on the EOS for the composition interpretation, the original density profile is more agnostic and the assumed temperature profile does not necessarily follow the  adiabatic one \citep[e.g.,][]{Morf2024}.  Figure~2 shows a selection of  inferred density profiles of Uranus and Neptune.  The shaded areas show the range of solutions from the empirical models presented by   \citet{Neuenschwander2022}. For comparison, we also present density profiles inferred using physical models (see figure caption for details). 
\par


Empirical models of Uranus and Neptune suggest that the interiors are non-adiabatic and that the planets do not consist of distinct layers. In some regions, the planetary interior is found to have composition gradients and/or boundary layers. The composition interpretation from empirical models also challenges the notion that the interiors of Uranus and Neptune are water-rich \citep{helled11,Neuenschwander2024} as solutions with a rock-rich interiors are also possible. \citep{helled11} inferred a metallicity of $\sim$76\% to $\sim$ 90\% for Uranus when the heavy elements are  represented by SiO$_2$ (rock) and H$_2$O (water), respectively. For Neptune, they inferred heavy-element masses of 15.5 M$_{\oplus}$ (Z=0.904) and 13.1 (Z=0.766) M$_{\oplus}$,  when the heavies were represented by H$_2$O and SiO$_2$, respectively.  \citep{helled11} (see section 3.1 for further discussion). Finally, if Uranus and Neptune have non-adiabatic deep interiors, they can be significantly hotter with central temperatures of the order of up to 50,000 K, in comparison to central temperatures of $\sim$ 5,000 K as inferred from adiabatic models.  

\citet{Movshovitz22} used empirical models to investigate how the density profiles of Uranus and Neptune can be further constrained with improved measurements of their gravitational moments. It was also found that a fraction of Uranus’ envelope is consistent with an adiabatic region of H-He with solar atmospheric abundances \citep{Movshovitz22}. Neptune's interior was found to consist of a relatively large amount of elements heavier than water \citep{Movshovitz22}. 
\par 

\cite{Neuenschwander2024} presented empirical models for Neptune and predicted the $J_6$ and $J_8$ values. They showed that Uranus is
expected to have a convective atmosphere/mantle on top of a
non-convective inner region which is stable. The transition
between the convective and non-convective region depends on the specific model and can vary between $\sim$40\% and $\sim$85\% of
Uranus’ radius. It was also shown that that a precise measurement of the MoI of Neptune, with relative uncertainties of approximately 0.1\% (1\% for Uranus), could help to constrain the planetary rotation rate and depth of the winds (see section 5).  
\cite{Neuenschwander2024} also used the empirical models to identify  non-adiabatic regions in Uranus’ interior. They inferred higher internal temperatures (a few $10^4$ K)  and higher bulk heavy-element abundances (by up to 1 M$_{\oplus}$) compared to standard adiabatic models. 
Finally, they predicted the higher-order gravitational coefficients ($J_6$ and $J_8$) for Uranus and Neptune based on various assumptions regarding the planetary rotation periods, wind depths, and uncertainties in lower-order harmonics. 

\begin{figure}[h]
    \centering
\includegraphics[width=0.65\paperwidth]{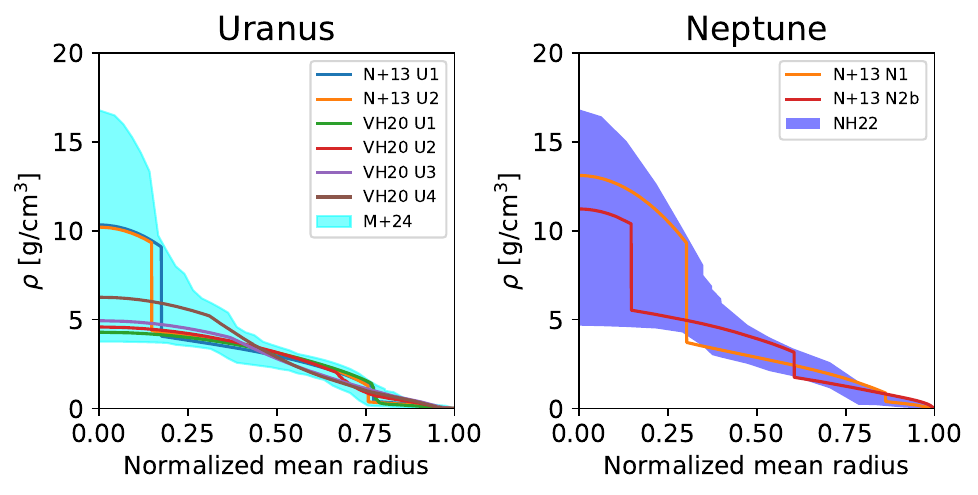}
    \caption{{\bf Top:} Density profiles of Uranus and Neptune. The shaded areas show the range of solutions from a set of empirical models from \citet{Morf2024}  (M+24) for Uranus and from \citet{Neuenschwander2022} (NH22) for Neptune. For Uranus we also show solutions from an adiabatic model (\citep{nettel13}, N+13, U1, U2) and evolution models (\citet{Vazan2020}, VH20, U1, U2, U3, U4). For Neptune we show two density profiles from  \citep{nettel13} (NN1, N2b). }
    \label{fig:rho}
\end{figure}

At the moment, using both modeling approaches (physical and empirical structure models), it is still not possible to determine the masses of rock and water in Uranus and Neptune (see section 3.1). In addition, it is also unclear whether water and rock are mixed in the deep interior. While the  traditional models assume that all the rocks (and possibly metals) are concentrated in a core, rock and water can be mixed, or partially mixed. Knowledge of the behavior of the phase diagram of a rock-water mixture is required in order to determine under what conditions rock and water are imicible  \citep[e.g.,][]{2022NatSR..1213055K,Darafeyeu_2024}.  
\par

\subsection{"Ice giants" or "rock giants"?}


Uranus and Neptune are commonly labeled as the "ice giants" due to the somewhat naive expectation that they contain large portions of water and other volatiles (methane, ammonia). This view  is based on the following assumptions: (1) The planets formed at large distances from the Sun, around 20-30 AU, where low temperatures in the solar nebula would favor the formation of water-rich solids that these planets could accrete; and (2) The presence of magnetic fields on Uranus and Neptune, which suggests that conductive materials like ionic water are in their deep interiors.
However, these assumptions are somewhat naive and based on misconceptions. 
First, it is now known that Pluto, despite its distance from the Sun, contains about 70\% rock. As a result, Kuiper belt objects bodies that formed in the outer part of the solar system are not necessarily icy. While oxygen is expected to be abundant in the solar nebula, currently, the oxygen-to-hydrogen ratios in the atmospheres of both Uranus and Neptune are not well-constrained. 
Second, other materials, such as compressed silicates mixed with hydrogen, could provide sufficient conductivity to generate magnetic fields. 
Thus, while the "ice giant" label is widely used, the true composition of Uranus and Neptune remains an open question, with potential scenarios ranging from water-rich worlds to planets with significant rocky interiors.

\begin{figure*}[h!]
    \centering
    \includegraphics[width = 0.65\textwidth]{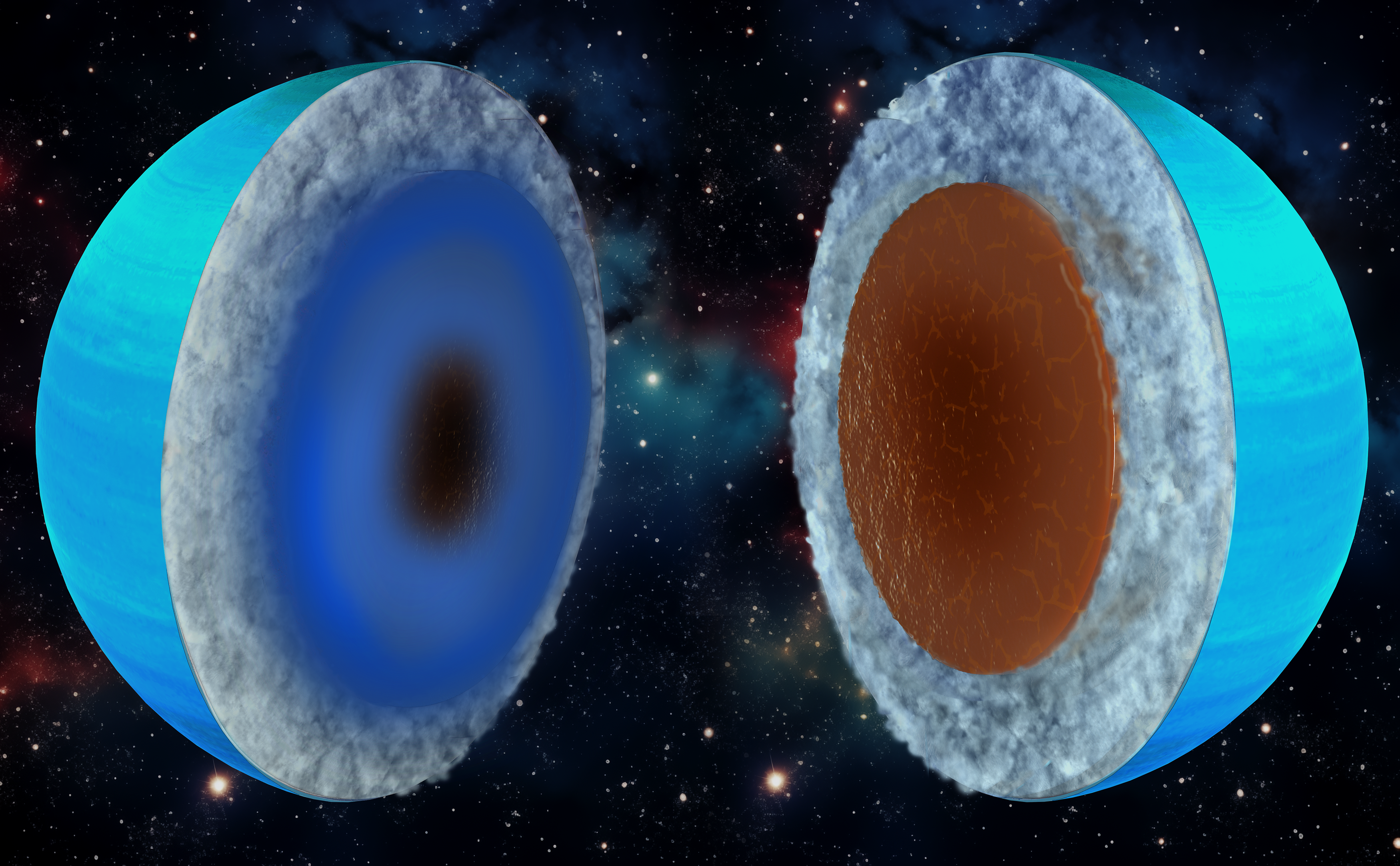}%
    \caption{Sketches of the two end-members of possible bulk compositions of Uranus and Neptune: a water-dominated interior (left) and a rock-dominated interior (right). The dark brown at the planetary center represents the core, which is expected to be composed of iron and rocks. The atmosphere is represented with gray. Note that the transition between different layers may be fuzzy and that currently it is still unknown whether the planetary interiors are differentiated with distinct compositional layers. Keck Institute for Space Studies / Chuck Carter}
    \label{fig:UranusRockIce}
\end{figure*}

Also, as discussed above, due to the limited available gravity data and the large measurement uncertainties, the bulk compositions of Uranus and Neptune remain largely uncertain. It is unclear whether Uranus and Neptune are predominantly water-dominated ("icy") or rock-dominated ("rocky"), as the inferred global water-to-rock ratio varies widely and is heavily dependent on the model assumptions.  Therefore, the name "ice giants"  for Uranus and Neptune might be inappropriate as it might not reflect their bulk compositions \citep{Helled2020,HelledFortney2020}. 
Figure 3 illustrates the two end-members of possible solutions for the planetary interiors: a water-dominated interior, "ice giants" (left) and a rock-dominated interior, "rock giants" (right). 
\par

Interior models of Uranus and Neptune  clearly show that the planets could actually be "rock giants" instead of "ice giants" with interiors composed of substantial amounts of silicates. 
\cite{Neuenschwander2024}  presented interior models of Uranus where the interior is rock-rich. 
Recently, \cite{Morf2024} modeled Uranus' interior with empirical models allowing  models of Uranus that differ with respect to the maximal number of materials allowing for several composition components to co-exist in the interior.  In the first case, where H-He is not allowed to exist in the deep interior (but only in the atmosphere) they find a rather high water abundance ($\sim$ 70\%) which also leads to lower central temperatures. On the other hand, if H-He can be mixed in the deeper interior, solutions where the planetary composition is rock-dominated could be found (with a water mass fraction of 30\% and a water-to-rock ratio of $\sim 0.6$ on average). The H-He mass fraction in the center was found to be upper-bounded by $\sim10\%$ of the core mass. Most of the models were found to be either purely convective with the exception of boundary layers, or only convective in the outermost region above $\sim 80 \%$ of Uranus' radius. Almost all of the models possess a region ranging between $\sim 75\%-90\%$ of Uranus' radius that is convective and consists of ionic water which could explain the generation of Uranus' magnetic field (see section 4). 
Recently, \cite{2024arXiv240312512M} suggested that Uranus and Neptune accreted significant fractions of planetesimals that are carbon-rich, similar to composition of comets. It was then found that the planets can be methane-rich and that rock-rich interiors are possible if the rock is mixed with hydrogen.  \citep{Morf2024} also derive internal structures that are rock-dominated if composition gradients exist and H-He is present in the planetary deep interior, as predicted by several formation models \citep{Helled2017, Valletta2022}.  

Finally, measurements of the atmospheric chemical composition can also help in constraining the planetary bulk composition. \cite{Teanby2020} showed that for Neptune although most existing CO profiles imply significant enrichment in oxygen, high O/H enrichment is inconsistent with D/H observations for a fully mixed and equilibrated Neptune. The measured CO and D/H can be consistent with a water-dominated internal structure only if Neptune is not fully mixed or if the interior is  rock-dominated and the measured CO in Neptune's troposphere is a result of external pollution. 
For Uranus we don't have measurements of the tropospheric CO. However, 
similar arguments could be applied to Uranus, which has similar  measured C/H and D/H enrichments. 


Overall, it is fair to say that at the moment it remains uncertain whether the bulk compositions of Uranus and Neptune  are primarily water-dominated or rock-dominated. Both scenarios are plausible, and further measurements are needed to determine the true composition of the planets. 


\section{Magnetic Fields}
Knowledge about the magnetic fields of planets can be used as an additional (and complementary) source of information for constraining their internal structures and heat transport.   
The existence of large-scale magnetic fields requires convective motions in a medium that is electrically conducting. While all the outer planets in the solar system have magnetic fields, their nature is rather different. The magnetic fields of Jupiter and Saturn are dipolar while the magnetic
fields of Uranus and Neptune are multipolar. 
The magnetic fields of both Uranus and Neptune also  have non-uniform structures, are tilted and asymmetric  \citep{con_ura,con_nep, holme}.

The magnetic fields of Uranus and Neptune were discovered and characterized during the respective Voyager 2 flybys in the mid 80s.  
Both Uranus and Neptune have magnetic fields that are tilted significantly relative to their rotation axes (Uranus: 59$^o$, Neptune: 47$^o$). This contrasts with Earth, Jupiter, and Saturn, whose fields are more aligned. In addition, Uranus and Neptune exhibit higher-order multipolar components, such as quadrupoles and octupoles. This results in complex variations in magnetic strength across the planet's surfaces. Due to their extreme tilts and offsets, the magnetospheres of these planets undergo dramatic changes as they rotate, unlike the more stable fields of Earth or Jupiter. These changes affect interactions with the solar wind and influence the structure and dynamics of the magnetospheres \citep{2018GeoRL..45.7320M}. 
\par

\begin{figure}
    \centering
    \includegraphics[width = 0.7\textwidth]{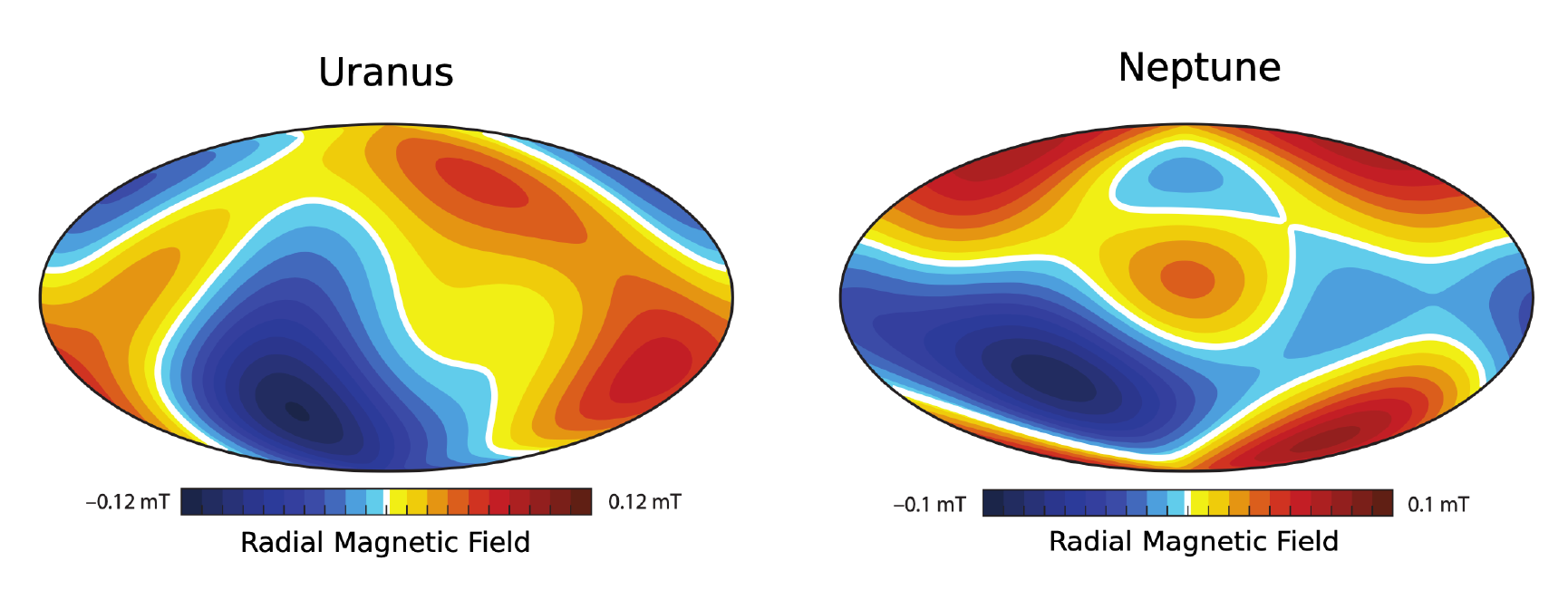}
    \caption{The radial component of Uranus’ (left) and Neptune’s (right) surface
magnetic field \cite{Tian2015PlanetaryDM}.}
    \label{fig:magfields}
\end{figure}

The magnetic fields in Uranus and Neptune are thought to be generated in a thin shell of ionic water (or other conductive material, e.g. ammonia, methane which becomes electrically conducting  at high pressures and temperatures). 
Convective motions within such a shell combined with the planetary rotation,  leads to the generation of sustainable magnetic fields (dynamo generation) \citep[e.g.,][]{redmer,krista2020}. 
While it is commonly assumed that the magnetic fields of Uranus and Neptune are generated due to ionic water, it should be noted that other mixtures (possibly even rocks mixed with H-He) could lead to a sufficiently high electrical conductivity. 
In addition, the existence of composition gradients in the planetary interior could also affect the density and conductivity profiles and therefore the characteristics of the magnetic fields. 
The exact locations of these shells are unknown but are expected to be located closer to the surface ($\sim$ 0.2--0.3 planetary radii below their surfaces). Figure \ref{fig:magfields} shows the radial component of the surface magnetic fields of Uranus and Neptune. 
\par

There is a clear connection between the existence of a magnetic field and the internal structure of planets. This interplay should be investigated further in future studies. In addition, accurate measurements of both the gravitational and magnetic fields are important to constrain the planetary interior and the nature of the magnetic fields of Uranus and Neptune. 



\section{Winds and rotation rates}

The atmospheres of Uranus and Neptune represent the outer layer of the  interior and correspond to the only region that can be observed from far. The composition and pressure-temperature ($P-T$) profile of these atmospheres establish boundary conditions for interior models and influence the cooling of the planets. Vertical and horizontal gradients in these profiles are closely connected to energy transport, shedding light on heat transfer mechanisms and energy loss to space. 
The compositions of the atmospheres are not well determined but are thought to consist of H-He with methane, ammonia and other hydrocarbons \citep[see][and references therein]{hueso}.

\begin{figure}
    \centering
\includegraphics[width=0.4\paperwidth]{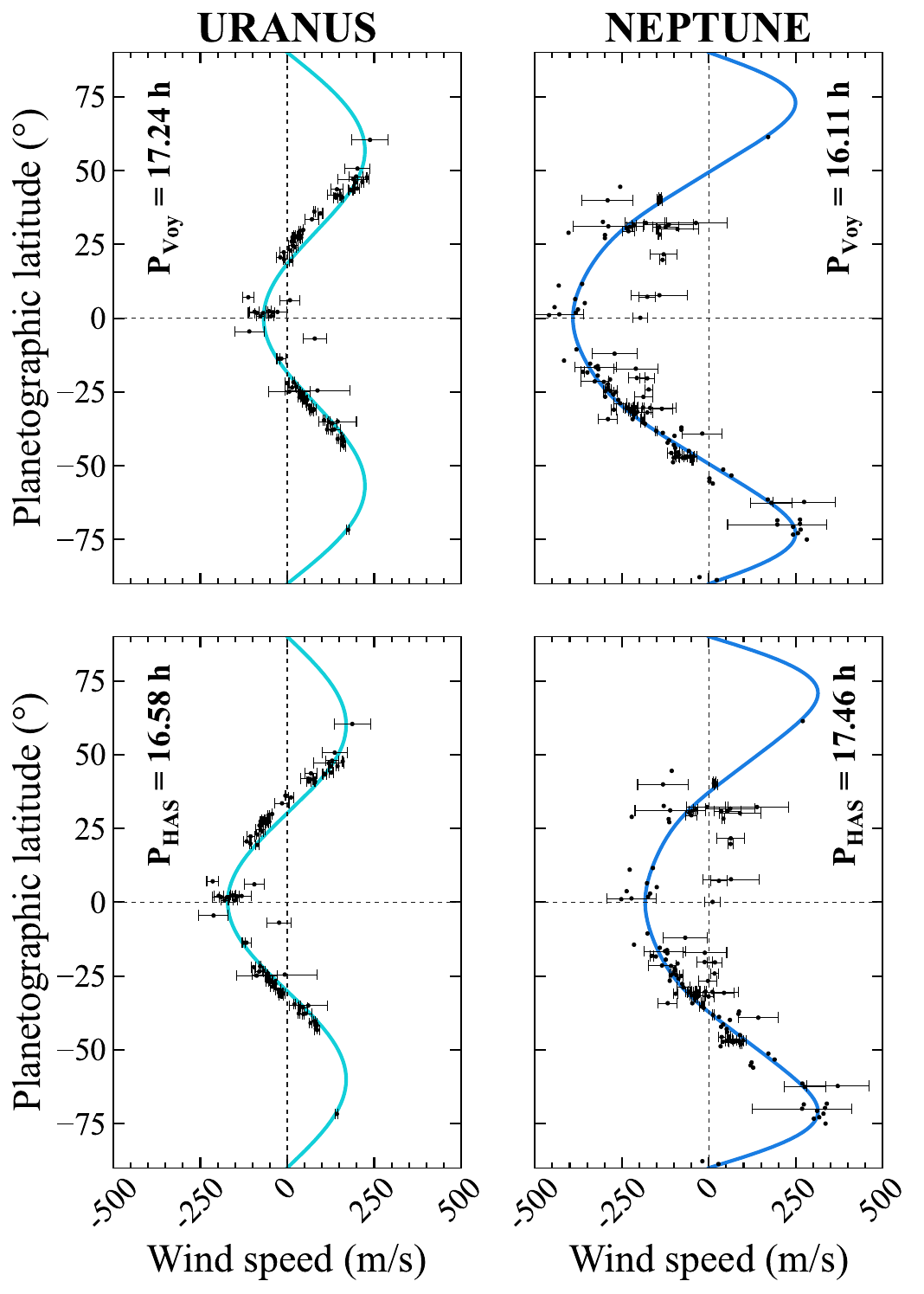}
    \caption{Zonal wind velocities in Uranus and Neptune as a function of latitude assuming the Voyager 2 ratio periods. {\bf Left}:  Zonal wind velocities for  Uranus for the Voyager (top) and modified rotation periods presented by \citep{Helled2010shape} (bottom). 
    {\bf Right}:  Zonal wind velocities for  Neptune for the Voyager (top) and modified rotation periods presented by \citep{Helled2010shape} (bottom). Note that the modified rotation periods were found by minimizing the wind speeds and
dynamical heights and that they lead to more similar atmosphere dynamics in the two planets.
The points with the error bars correspond to measurements and the lines to fits to the wind velocities inferred  by \citet{sromovsky} and \citet{french_wind} for Uranus and Neptune, respectively. 
}
    \label{fig:winds}
\end{figure}

Both Uranus and Neptune have fast surface zonal winds with velocities reaching up to  200 ms$^{-1}$ and 400 ms$^{-1}$, respectively. The measured wind velocities are shown in Figure \ref{fig:winds}. Note, however, that the velocities are given relative to the assumed bulk rotations, which are in fact not well known for both planets. 
The Voyager 2 rotation period of Uranus and Neptune based on radio data are 17.24 h and 16.11 h, respectively. However, we now know, thanks to measurements from the Cassini mission at Saturn, that Voyager's measured periodicity may not represent the rotation rate of the deep interior. \citet{helled_shape} suggested modified rotation periods for the planets based on the minimization of the wind velocities (and the dynamical heights of their surfaces). The inferred rotation periods were found to be 16.58 h for Uranus and 17.46 h for Neptune. 
With these rotation periods the wind velocities are less extreme and more symmetric between east and west in both planets (see Figure \ref{fig:winds}). 
The planetary rotation periods affect the density distribution within the planets and their oblatenesses  and therefore strongly influence the inferred internal structure and
composition of the planets. For example, for  Uranus, interior models that use a faster rotation period are associated with hotter interiors, a higher 
heavy-element mass, a lower water-
to-rock ratio and a larger convective
region \citep{Neuenschwander2024}. Also, interestingly,  the modified rotation periods suggested by \citet{helled_shape} imply that the bulk compositions of Uranus and Neptune are rather different \citep[see][for details]{nettel13}.   It is therefore clear that determining the rotation periods of Uranus and Neptune is crucial. 
\par 

The zonal winds on Uranus and Neptune are measured to  be rather symmetrical relative to the equator, exhibiting retrograde motion at central latitudes and prograde motion at higher latitudes. It is unclear, however, how deep the winds penetrate into the planetary interior.  Currently, it remains uncertain whether the winds extend deep into the planetary interior or are confined to shallower (weather) layers. The mechanisms behind the decay of these winds are also not fully understood. 
\citet{kaspi2013} constrained the depth of the winds in Uranus and Neptune by analyzing the gravitational harmonics of the planets, which encode deviations in the planets' gravitational fields caused by density variations induced by atmospheric winds. They estimated that the winds penetrate to depths of approximately 1,000 km on both planets. These findings suggest that the winds are confined to the outer atmospheric layers, with their influence diminishing in the denser interior regions. 
Recent studies which use constraints from Ohmic dissipation limits and the available gravitational harmonics confirmed that the winds in Uranus and Neptune are expected to dissipate rapidly within the shallow layers of Uranus and Neptune \citep{soyuer2020, soyuer2023}.

\section{The Origin of Uranus and Neptune}

The formation of Uranus and Neptune has been a long-standing challenge for planet formation theory. 
In the framework of the standard core-accretion model \citep[e.g.,][]{Pollack1996}, a heavy-element core is formed first by the accretion of solids, which can range in size from cm (pebbles) to km (planetesimals). 
Once the core reaches a few M$_{\oplus}$, it is able to bind a tenuous atmosphere (gaseous envelope), composed primarily of H-He. 
The protoplanet then continues to grow by accreting both solids and gas from the protoplanetary disk. Once the mass of the envelope is comparable to the core's mass, the self-gravity of the envelope becomes strong enough to contract rapidly,  triggering rapid gas accretion (runaway gas accretion). 
Since structure models of Uranus and Neptune suggest that the planets consist of H-He atmospheres of only a few M$_{\oplus}$ (unlike Jupiter and Saturn which consist of much higher fractions of H-He), it is reasonable to assume that they grew relatively slowly and never reached the conditions that lead to rapid (and significant) gas accretion. 
Indeed, it was recently suggested that giant planet formation can take several million years with runaway gas accretion being delayed due to a continuous planetesimal accretion \citep{Helled2023}. 
The delay of H-He accretion by a few million years can explain why Uranus and Neptune are heavy-element dominated in composition and have not become gas giant planets. 
\par 

While it is clear that the formation process of Uranus and Neptune must have been long (of the order of a few million years), it is still challenging for theoretical models to form Uranus and Neptune in their current observed distances. This is because at such large distances from the sun, the solid surface density in the disk is expected to be low. Traditional core accretion models suggest that the formation timescales of Uranus and Neptune significantly exceed the expected lifetimes of protoplanetary disks \citep[e.g.,][]{2015A&A...576A..52R}.  It was therefore suggested by several authors that Uranus and Neptune have formed at radial distances of 5-15 AU and then migrated outwards \citep[e.g.,][]{Tsiganis2005}. 
Other ways to reduce the formation timescales of the icy giants include  high-accretion rates as a result of pebble accretion \citep[e.g.,][]{2014A&A...572A.107L} or a dynamically cold planetesimal disk \citep[e.g.,][]{2004ARA&A..42..549G} . 
Another suggested formation  path  is collision and merging of a few low-mass protoplanets which accreted from a population of planetary embryos which also lead to shorter formation timescales \citep[e.g.,][and references therein]{chau_impact}.  
However, a key remaining  challenge  for planet formation  models    is to reproduce the final masses and compositions of Uranus and Neptune when considering the accretion of both the solids (heavy elements) and gas (H-He). This becomes even more challenging when searching for formation conditions that can satisfy simultaneously the constraints of both Uranus and Neptune and their mutual growth.  

\begin{figure}[h]
	\centering
	\includegraphics[width=0.55\linewidth]{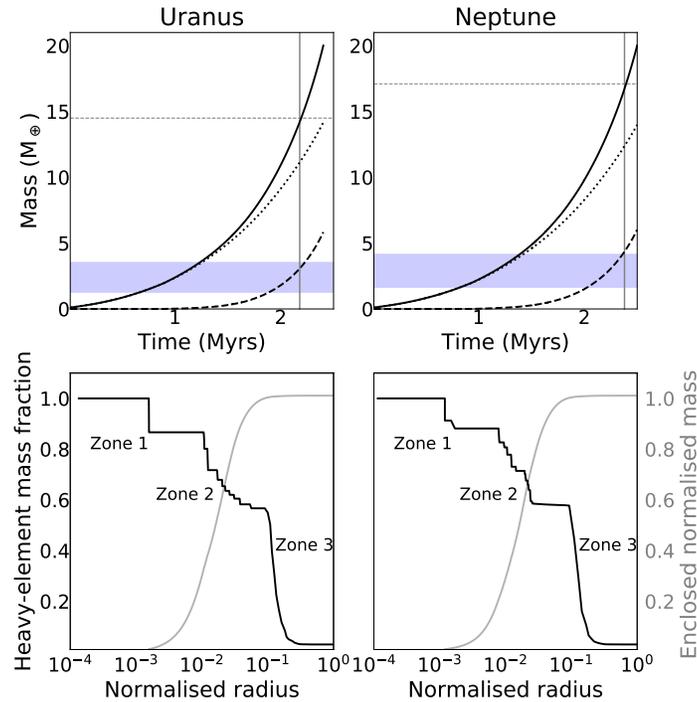}
	\caption{
	The formation of Uranus (left) and Neptune (right). 
	\textbf{Top panels:}
	Planetary mass as a function of time. The solid, dashed, and dotted lines show the total planetary mass, the H-He mass, and the mass of heavy elements, respectively. 
	The gray horizontal lines indicate the mass of Uranus and Neptune, while the vertical gray line shows the required disk's dissipation time.
	The blue region indicates the H-He mass in Uranus and Neptune as inferred from interior models. 
    \textbf{Bottom panels:} internal profile when the disk dissipates. The black line shows the heavy-element mass fraction vs.~normalised radius. The gray line shows the  enclosed mass (Figure from \cite{Valletta2022}).
	}
	\label{Uranus-Neptune}
\end{figure}

Recently, \cite{Valletta2022} investigated whether Uranus and Neptune could have formed at their current locations via pebble accretion . It was shown that Uranus and Neptune  can grow efficiently and reach their current masses within the lifetime of the solar nebula even at their current radial distances and have masses and compositions consistent with interior models. However, the probability of forming Uranus and Neptune and fit their observed properties was found to be low. The inferred formation probability was found to strongly depend on the assumed H-He mass fractions in Uranus and Neptune.  
Figure 6 shows examples of simulations that led to the in-situ formation of Uranus and Neptune. These successful simulations result in the correct final planetary mass as well as the H-He mass. The top panels show the planetary growth as a function of time for Uranus (left) and Neptune (right). 
The bottom panels  show the post-formation heavy-element distributions from these simulations. The heavy-element mass fraction is found to decrease  towards the planetary surface. Also, both planets are predicted to have small pure heavy-element cores ($\sim$ M$_{\oplus}$) and composition gradients. These formation models predict that the deep interior also includes H-He  and that the outermost part of the planets (i.e., the atmospheres) are dominated by H-He. These predictions are consistent with several interior models of Uranus and Neptune as discussed in section 3. 
Interestingly, the study of \cite{Valletta2022} also demonstrated that in many cases the formation path led to proto-Uranus and proto-Neptune that are missing $\sim$ 1-3  M$_{\oplus}$ of heavy elements. This missing heavy-element mass could be added  after their formation via planetesimal accretion during planetary migration \citep{2024AJ....168...64Z} and/or giant impacts (see section 6.1).
The results by \citet{Valletta2022}  suggest that formation by pebble accretion can match the low mass and composition of the ice giants, supporting the viability of this model over the traditional core accretion scenario. At the same time, it is clear that the formation of the planets require rather specific formation conditions and that from a statistical perspective, the formation of the planet is rather challenging.  \citet{Eriksson2023} investigated the  concurrent formation of Uranus and Neptune accounting for both pebble and planetesimal accretion. They concluded that it is unlikely for Uranus and Neptune to
have formed {\it in-situ} where the main  challenge is to keeping the
H-He mass fractions below 20\% and keeping the planetary masses
similar.  They found that Uranus and Neptune analogues can form when
the orbital distance between the planets is small and when the outer embryo is 10 times more massive
than the inner embryo. Their study highlights the low probability of forming Uranus and Neptune, and the  complexity of modeling the formation of the planets when accounting the disk evolution and the accretion rates properly.   
Overall, despite the variety of proposed models for the origin of the ice giants, there is still no satisfactory formation model for Uranus and Neptune.

\subsection{Giant impacts on Uranus and Neptune}
The unique characteristics of Uranus and Neptune, particularly Uranus's extreme axial tilt and the similar yet distinct internal structures of both planets, are often attributed to giant impacts shortly after their formation, when impacts were still very common in the young solar system. 
While other scenarios to explain Uranus's axial tilt have been suggested \citep[e.g.,][]{2010ApJ...712L..44B,2022A&A...668A.108S}, several 
formation models suggest that both planets experienced at least one major collision \citep[e.g.,][]{formation4,chau_impact}. 
\citet{2012ApJ...759L..32P} suggested that impacts with different geometries could explain the observed axial tilt, satellite system, heat fluxes and inferred moment of inertia of the planets.

\par 

Impact simulations indeed confirm that in case of Uranus an oblique collision with a massive object (2-3 M$_{\oplus}$) can tilt the planet on its side \citep[e.g.,][]{kegerreis,Reinhardt2020}. Such an impact also ejects enough material that forms a  cirumplanetary disk that allows the formation of Uranus’ regular moons. In the case of an oblique impact, the deeper interior of the planet is unaffected. 
On the other hand, a more head-on collision\footnote{Note that simulations suggest that head-on collisions are less probable and most collisions occur at an oblique incidence.} on young Neptune could have mixed its interior, leading to a convective interior which explains its higher heat flux compared to Uranus, and its higher inferred  moment of inertia. In addition, in the case of a head-on collision a circumplanetary disk is not formed which is consistent with the lack of regular moons around Neptune and the fact that Neptune's largest moon Triton is an irregular satellite and is probably a captured object.
 It is therefore possible that both Uranus and Neptune suffered from giant impacts with Uranus experiencing an oblique impact while Neptune a more "head-on" collision (see \citep{Reinhardt2020} and Fig.~5 for details).  While this scenario is possible, the origin of the differences between Uranus and Neptune, in particular, their satellite system are unknown and further work on giant impacts on Uranus and Neptune (and their link to the internal structure)  are clearly desirable.

\begin{figure}[h]
	\centering
	\includegraphics[width=0.605\linewidth]{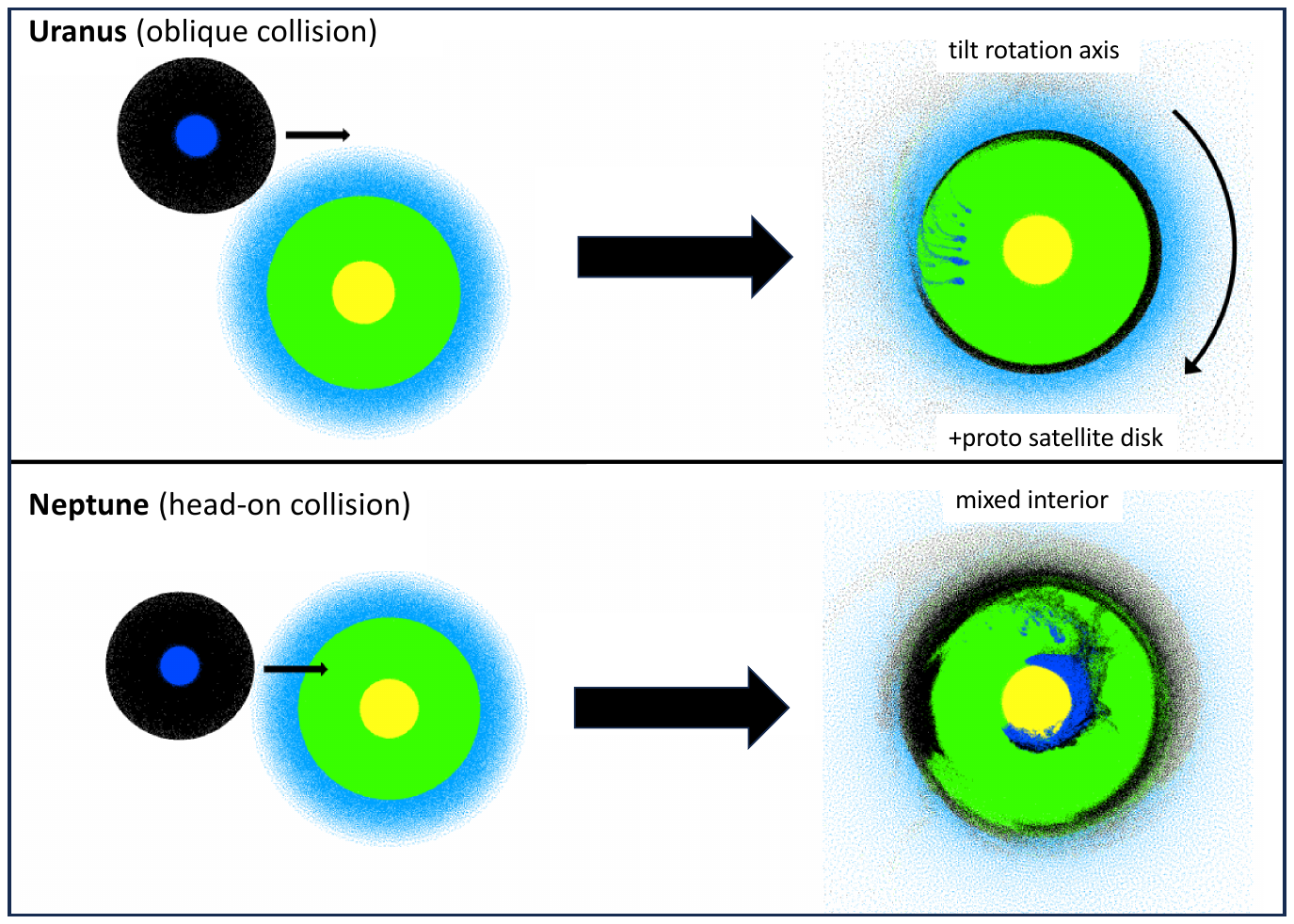}
	\caption{A sketch presenting the idea of the role of giant impacts in explaining the dichotomy
between Uranus and Neptune (not to scale). An oblique giant impact on Uranus could tilt
its spin axis significantly and eject enough material to form a disk and the regular satellites
while keeping the body stratified. For Neptune, an almost head-on collision might deposit
energy deep inside, mixing its interior resulting in a thermal profile that is close to adiabatic
explaining the fast cooling (figure based on \citet{Reinhardt2020}).}
	\label{Uranus-Neptune}
\end{figure}


\newpage
\section{Summary \& Outlook}
 In this chapter we summarized the current knowledge of the  internal structure and formation history of Uranus and Neptune.  As discussed above, despite their classification as "ice giants," it is in fact unclear whether Uranus and Neptune are indeed "icy" and these planets remain the least explored and least understood planets in the solar system.  
 While we know that both planets have interiors that are dominated by heavy-elements, the bulk compositions, in particular the rock-to-water ratio and the way the different materials are distributed within the planets remain poorly  constrained.  It is also still unclear how different the two planets are from each other. We next  briefly introduced the complex magnetic fields of Uranus and Neptune and their atmospheres which have strong zonal winds. Both of these aspects are still being explored. 
 Finally, we discussed the challenge in constraining the formation histories of Uranus and Neptune and the potential role of giant impacts in their early evolution.  
 \par

 Finally, it is clear that Uranus and Neptune should have dedicated missions that can reveal some of their mysteries and provide measurements with lower uncertainties. Indeed, the recent planetary decadal survey \citep{NRC2022} indicated that missions to Uranus and Neptune are of highest priority when it comes to the exploration of our solar system. 
 There is a clear need to  advance our understanding of the outer solar system and address key scientific questions regarding these planets' formation, evolution, and internal structure, which differ significantly from the gas giants, Jupiter and Saturn. Specifically, such future missions would allow for in-depth studies of their complex atmospheres and interiors, and unique magnetic fields, offering insights into the processes that govern planetary systems both within and beyond our solar system. A dedicated mission to each planet will also allow us to better understand the differences between the two planets and their system and therefore on the diversity of intermediate-mass planets in more general terms. 
\par 

Despite the many remaining open questions, the future of studying the interiors of Uranus and Neptune looks bright  with the expected advancements in both observations and theoretical/numerical models.  In addition to the planned future space mission, refining our understanding of the equation of state for various elements and their interactions, while integrating  diverse datasets (gravity fields, magnetic fields, atmospheric compositions, etc.), is crucial for refining our grasp of planetary interiors. 
Technological advancements in high-pressure physics experiments  and supercomputing will allow for an improved understanding of materials at extreme conditions. 
 
 We are convinced that integrating information from different future measurements and theoretical investigations can deepen our understanding of the  formation, evolution, and interiors of Uranus and Neptune. 


\begin{ack}[Acknowledgments]
W We acknowledge support from SNSF grant \texttt{\detokenize{200020_188460}} and the National Centre for Competence in Research ‘PlanetS’ supported by SNSF. This work is supported in part by the W.M. Keck Institute for Space Studies. 
\end{ack}


\bibliographystyle{Harvard}
\bibliography{reference}

\begin{thebibliography*}{53}
\providecommand{\bibtype}[1]{}
\providecommand{\natexlab}[1]{#1}
{\catcode`\|=0\catcode`\#=12\catcode`\@=11\catcode`\\=12
|immediate|write|@auxout{\expandafter\ifx\csname natexlab\endcsname\relax\gdef\natexlab#1{#1}\fi}}
\renewcommand{\url}[1]{{\tt #1}}
\providecommand{\urlprefix}{URL }
\expandafter\ifx\csname urlstyle\endcsname\relax
  \providecommand{\doi}[1]{doi:\discretionary{}{}{}#1}\else
  \providecommand{\doi}{doi:\discretionary{}{}{}\begingroup \urlstyle{rm}\Url}\fi
\providecommand{\bibinfo}[2]{#2}
\providecommand{\eprint}[2][]{\url{#2}}

\bibtype{Article}%
\bibitem[{Bailey} and {Stevenson}(2021)]{Bailey2021}
\bibinfo{author}{{Bailey} E} and  \bibinfo{author}{{Stevenson} DJ} (\bibinfo{year}{2021}), \bibinfo{month}{Apr.}
\bibinfo{title}{{Thermodynamically Governed Interior Models of Uranus and Neptune}} \bibinfo{volume}{2} (\bibinfo{number}{2}), \bibinfo{eid}{64}. \bibinfo{doi}{\doi{10.3847/PSJ/abd1e0}}.
\eprint{2012.04166}.

\bibtype{Article}%
\bibitem[{Bou{\'e}} and {Laskar}(2010)]{2010ApJ...712L..44B}
\bibinfo{author}{{Bou{\'e}} G} and  \bibinfo{author}{{Laskar} J} (\bibinfo{year}{2010}), \bibinfo{month}{Mar.}
\bibinfo{title}{{A Collisionless Scenario for Uranus Tilting}}.
\bibinfo{journal}{{\em \apjl}} \bibinfo{volume}{712} (\bibinfo{number}{1}): \bibinfo{pages}{L44--L47}. \bibinfo{doi}{\doi{10.1088/2041-8205/712/1/L44}}.
\eprint{0912.0181}.

\bibtype{Article}%
\bibitem[{Chau} et al.(2021)]{chau_impact}
\bibinfo{author}{{Chau} A}, \bibinfo{author}{{Reinhardt} C}, \bibinfo{author}{{Izidoro} A}, \bibinfo{author}{{Stadel} J} and  \bibinfo{author}{{Helled} R} (\bibinfo{year}{2021}), \bibinfo{month}{Apr.}
\bibinfo{title}{{Could Uranus and Neptune form by collisions of planetary embryos?}}
\bibinfo{journal}{{\em Monthly Notices of the Royal Astronomical Society}} \bibinfo{volume}{502} (\bibinfo{number}{2}): \bibinfo{pages}{1647--1660}. \bibinfo{doi}{\doi{10.1093/mnras/staa4021}}.
\eprint{2009.10100}.

\bibtype{Article}%
\bibitem[{Connerney} et al.(1987)]{con_ura}
\bibinfo{author}{{Connerney} JEP}, \bibinfo{author}{{Acuna} MH} and  \bibinfo{author}{{Ness} NF} (\bibinfo{year}{1987}), \bibinfo{month}{Dec.}
\bibinfo{title}{{The magnetic field of Uranus}}.
\bibinfo{journal}{{\em Journal of Geophysics Research}} \bibinfo{volume}{92} (\bibinfo{number}{A13}): \bibinfo{pages}{15329--15336}. \bibinfo{doi}{\doi{10.1029/JA092iA13p15329}}.

\bibtype{Article}%
\bibitem[{Connerney} et al.(1991)]{con_nep}
\bibinfo{author}{{Connerney} JEP}, \bibinfo{author}{{Acuna} MH} and  \bibinfo{author}{{Ness} NF} (\bibinfo{year}{1991}), \bibinfo{month}{Oct.}
\bibinfo{title}{{The magnetic field of Neptune}}.
\bibinfo{journal}{{\em Journal of Geophysics Research}} \bibinfo{volume}{96}: \bibinfo{pages}{19023--19042}. \bibinfo{doi}{\doi{10.1029/91JA01165}}.

\bibtype{Article}%
\bibitem[Darafeyeu et al.(2024)]{Darafeyeu_2024}
\bibinfo{author}{Darafeyeu V}, \bibinfo{author}{Rimle S}, \bibinfo{author}{Mazzola G} and  \bibinfo{author}{Helled R} (\bibinfo{year}{2024}), \bibinfo{month}{nov}.
\bibinfo{title}{The linear mixing approximation in silica–water mixtures at planetary conditions}.
\bibinfo{journal}{{\em The Astrophysical Journal}} \bibinfo{volume}{975} (\bibinfo{number}{2}): \bibinfo{pages}{255}. \bibinfo{doi}{\doi{10.3847/1538-4357/ad7e29}}.
\bibinfo{url}{\url{https://dx.doi.org/10.3847/1538-4357/ad7e29}}.

\bibtype{Article}%
\bibitem[{Eriksson} et al.(2023)]{Eriksson2023}
\bibinfo{author}{{Eriksson} LEJ}, \bibinfo{author}{{Mol Lous} MAS}, \bibinfo{author}{{Shibata} S} and  \bibinfo{author}{{Helled} R} (\bibinfo{year}{2023}), \bibinfo{month}{Dec.}
\bibinfo{title}{{Can Uranus and Neptune form concurrently via pebble, gas, and planetesimal accretion?}}
\bibinfo{journal}{{\em \mnras}} \bibinfo{volume}{526} (\bibinfo{number}{4}): \bibinfo{pages}{4860--4876}. \bibinfo{doi}{\doi{10.1093/mnras/stad3007}}.
\eprint{2310.00075}.

\bibtype{Article}%
\bibitem[French et al.(1998)]{french_wind}
\bibinfo{author}{French RG}, \bibinfo{author}{McGhee CA} and  \bibinfo{author}{Sicardy B} (\bibinfo{year}{1998}).
\bibinfo{title}{Neptune's stratospheric winds from three central flash occultations}.
\bibinfo{journal}{{\em Icarus}} \bibinfo{volume}{136} (\bibinfo{number}{1}): \bibinfo{pages}{27 -- 49}.
ISSN \bibinfo{issn}{0019-1035}. \bibinfo{doi}{\doi{https://doi.org/10.1006/icar.1998.6001}}.
\bibinfo{url}{\url{http://www.sciencedirect.com/science/article/pii/S0019103598960018}}.

\bibtype{Article}%
\bibitem[{Goldreich} et al.(2004)]{2004ARA&A..42..549G}
\bibinfo{author}{{Goldreich} P}, \bibinfo{author}{{Lithwick} Y} and  \bibinfo{author}{{Sari} R} (\bibinfo{year}{2004}), \bibinfo{month}{Sep.}
\bibinfo{title}{{Planet Formation by Coagulation: A Focus on Uranus and Neptune}}.
\bibinfo{journal}{{\em Annual Review of Astronomy and Astrophysics}} \bibinfo{volume}{42} (\bibinfo{number}{1}): \bibinfo{pages}{549--601}. \bibinfo{doi}{\doi{10.1146/annurev.astro.42.053102.134004}}.
\eprint{astro-ph/0405215}.

\bibtype{Article}%
\bibitem[{Helled}(2023)]{Helled2023}
\bibinfo{author}{{Helled} R} (\bibinfo{year}{2023}), \bibinfo{month}{Jul.}
\bibinfo{title}{{The mass of gas giant planets: Is Saturn a failed gas giant?}}
\bibinfo{journal}{{\em \aap}} \bibinfo{volume}{675}, \bibinfo{eid}{L8}. \bibinfo{doi}{\doi{10.1051/0004-6361/202346850}}.
\eprint{2306.14740}.

\bibtype{Article}%
\bibitem[{Helled} and {Fortney}(2020)]{HelledFortney2020}
\bibinfo{author}{{Helled} R} and  \bibinfo{author}{{Fortney} JJ} (\bibinfo{year}{2020}), \bibinfo{month}{Dec.}
\bibinfo{title}{{The interiors of Uranus and Neptune: current understanding and open questions}}.
\bibinfo{journal}{{\em Philosophical Transactions of the Royal Society of London Series A}} \bibinfo{volume}{378} (\bibinfo{number}{2187}), \bibinfo{eid}{20190474}. \bibinfo{doi}{\doi{10.1098/rsta.2019.0474}}.
\eprint{2007.10783}.

\bibtype{Article}%
\bibitem[{Helled} and {Howard}(2024)]{HelledHoward2024}
\bibinfo{author}{{Helled} R} and  \bibinfo{author}{{Howard} S} (\bibinfo{year}{2024}), \bibinfo{month}{Jul.}
\bibinfo{title}{{Giant planet interiors and atmospheres}}.
\bibinfo{journal}{{\em arXiv e-prints}} , \bibinfo{eid}{arXiv:2407.05853}\bibinfo{doi}{\doi{10.48550/arXiv.2407.05853}}.
\eprint{2407.05853}.

\bibtype{Article}%
\bibitem[{Helled} and {Stevenson}(2017)]{Helled2017}
\bibinfo{author}{{Helled} R} and  \bibinfo{author}{{Stevenson} D} (\bibinfo{year}{2017}), \bibinfo{month}{May}.
\bibinfo{title}{{The Fuzziness of Giant Planets{\textquoteright} Cores}}.
\bibinfo{journal}{{\em \apjl}} \bibinfo{volume}{840} (\bibinfo{number}{1}), \bibinfo{eid}{L4}. \bibinfo{doi}{\doi{10.3847/2041-8213/aa6d08}}.
\eprint{1704.01299}.

\bibtype{Article}%
\bibitem[{Helled} et al.(2010{\natexlab{a}})]{Helled2010shape}
\bibinfo{author}{{Helled} R}, \bibinfo{author}{{Anderson} JD} and  \bibinfo{author}{{Schubert} G} (\bibinfo{year}{2010}{\natexlab{a}}), \bibinfo{month}{Nov.}
\bibinfo{title}{{Uranus and Neptune: Shape and rotation}}.
\bibinfo{journal}{{\em \icarus}} \bibinfo{volume}{210} (\bibinfo{number}{1}): \bibinfo{pages}{446--454}. \bibinfo{doi}{\doi{10.1016/j.icarus.2010.06.037}}.
\eprint{1006.3840}.

\bibtype{Article}%
\bibitem[{Helled} et al.(2010{\natexlab{b}})]{helled_shape}
\bibinfo{author}{{Helled} R}, \bibinfo{author}{{Anderson} JD} and  \bibinfo{author}{{Schubert} G} (\bibinfo{year}{2010}{\natexlab{b}}), \bibinfo{month}{Nov.}
\bibinfo{title}{{Uranus and Neptune: Shape and rotation}}.
\bibinfo{journal}{{\em \icarus}} \bibinfo{volume}{210} (\bibinfo{number}{1}): \bibinfo{pages}{446--454}. \bibinfo{doi}{\doi{10.1016/j.icarus.2010.06.037}}.
\eprint{1006.3840}.

\bibtype{Article}%
\bibitem[{Helled} et al.(2011)]{helled11}
\bibinfo{author}{{Helled} R}, \bibinfo{author}{{Anderson} JD}, \bibinfo{author}{{Podolak} M} and  \bibinfo{author}{{Schubert} G} (\bibinfo{year}{2011}), \bibinfo{month}{Jan.}
\bibinfo{title}{{Interior Models of Uranus and Neptune}}.
\bibinfo{journal}{{\em \apj}} \bibinfo{volume}{726} (\bibinfo{number}{1}), \bibinfo{eid}{15}. \bibinfo{doi}{\doi{10.1088/0004-637X/726/1/15}}.
\eprint{1010.5546}.

\bibtype{Article}%
\bibitem[{Helled} et al.(2020)]{Helled2020}
\bibinfo{author}{{Helled} R}, \bibinfo{author}{{Nettelmann} N} and  \bibinfo{author}{{Guillot} T} (\bibinfo{year}{2020}), \bibinfo{month}{Mar.}
\bibinfo{title}{{Uranus and Neptune: Origin, Evolution and Internal Structure}}.
\bibinfo{journal}{{\em \ssr}} \bibinfo{volume}{216} (\bibinfo{number}{3}), \bibinfo{eid}{38}. \bibinfo{doi}{\doi{10.1007/s11214-020-00660-3}}.
\eprint{1909.04891}.

\bibtype{Article}%
\bibitem[Holme and Bloxham(1996)]{holme}
\bibinfo{author}{Holme R} and  \bibinfo{author}{Bloxham J} (\bibinfo{year}{1996}).
\bibinfo{title}{The magnetic fields of uranus and neptune: Methods and models}.
\bibinfo{journal}{{\em Journal of Geophysical Research: Planets}} \bibinfo{volume}{101} (\bibinfo{number}{E1}): \bibinfo{pages}{2177--2200}. \bibinfo{doi}{\doi{10.1029/95JE03437}}.
\eprint{https://agupubs.onlinelibrary.wiley.com/doi/pdf/10.1029/95JE03437}, \bibinfo{url}{\url{https://agupubs.onlinelibrary.wiley.com/doi/abs/10.1029/95JE03437}}.

\bibtype{Article}%
\bibitem[{Hueso} et al.(2020)]{hueso}
\bibinfo{author}{{Hueso} R}, \bibinfo{author}{{Guillot} T} and  \bibinfo{author}{{S{\'a}nchez-Lavega} A} (\bibinfo{year}{2020}), \bibinfo{month}{Dec.}
\bibinfo{title}{{Convective storms and atmospheric vertical structure in Uranus and Neptune}}.
\bibinfo{journal}{{\em Philosophical Transactions of the Royal Society of London Series A}} \bibinfo{volume}{378} (\bibinfo{number}{2187}), \bibinfo{eid}{20190476}. \bibinfo{doi}{\doi{10.1098/rsta.2019.0476}}.
\eprint{2111.15494}.

\bibtype{Article}%
\bibitem[{Izidoro} et al.(2015)]{formation4}
\bibinfo{author}{{Izidoro} A}, \bibinfo{author}{{Morbidelli} A}, \bibinfo{author}{{Raymond} SN}, \bibinfo{author}{{Hersant} F} and  \bibinfo{author}{{Pierens} A} (\bibinfo{year}{2015}), \bibinfo{month}{Oct.}
\bibinfo{title}{{Accretion of Uranus and Neptune from inward-migrating planetary embryos blocked by Jupiter and Saturn}}.
\bibinfo{journal}{{\em Astronomy and Astrophysics}} \bibinfo{volume}{582}, \bibinfo{eid}{A99}. \bibinfo{doi}{\doi{10.1051/0004-6361/201425525}}.
\eprint{1506.03029}.

\bibtype{Article}%
\bibitem[{Kaspi} et al.(2013)]{kaspi2013}
\bibinfo{author}{{Kaspi} Y}, \bibinfo{author}{{Showman} AP}, \bibinfo{author}{{Hubbard} WB}, \bibinfo{author}{{Aharonson} O} and  \bibinfo{author}{{Helled} R} (\bibinfo{year}{2013}), \bibinfo{month}{May}.
\bibinfo{title}{{Atmospheric confinement of jet streams on Uranus and Neptune}}.
\bibinfo{journal}{{\em Nature Astronomy Letters}} \bibinfo{volume}{497} (\bibinfo{number}{7449}): \bibinfo{pages}{344--347}. \bibinfo{doi}{\doi{10.1038/nature12131}}.

\bibtype{Article}%
\bibitem[{Kegerreis} et al.(2018)]{kegerreis}
\bibinfo{author}{{Kegerreis} JA}, \bibinfo{author}{{Teodoro} LFA}, \bibinfo{author}{{Eke} VR}, \bibinfo{author}{{Massey} RJ}, \bibinfo{author}{{Catling} DC}, \bibinfo{author}{{Fryer} CL}, \bibinfo{author}{{Korycansky} DG}, \bibinfo{author}{{Warren} MS} and  \bibinfo{author}{{Zahnle} KJ} (\bibinfo{year}{2018}), \bibinfo{month}{Jul.}
\bibinfo{title}{{Consequences of Giant Impacts on Early Uranus for Rotation, Internal Structure, Debris, and Atmospheric Erosion}}.
\bibinfo{journal}{{\em Astrophysical Journal}} \bibinfo{volume}{861} (\bibinfo{number}{1}), \bibinfo{eid}{52}. \bibinfo{doi}{\doi{10.3847/1538-4357/aac725}}.
\eprint{1803.07083}.

\bibtype{Article}%
\bibitem[{Kova{\v{c}}evi{\'c}} et al.(2022)]{2022NatSR..1213055K}
\bibinfo{author}{{Kova{\v{c}}evi{\'c}} T}, \bibinfo{author}{{Gonz{\'a}lez-Cataldo} F}, \bibinfo{author}{{Stewart} ST} and  \bibinfo{author}{{Militzer} B} (\bibinfo{year}{2022}), \bibinfo{month}{Jul.}
\bibinfo{title}{{Miscibility of rock and ice in the interiors of water worlds}}.
\bibinfo{journal}{{\em Scientific Reports}} \bibinfo{volume}{12}, \bibinfo{eid}{13055}. \bibinfo{doi}{\doi{10.1038/s41598-022-16816-w}}.

\bibtype{Article}%
\bibitem[{Lambrechts} and {Johansen}(2014)]{2014A&A...572A.107L}
\bibinfo{author}{{Lambrechts} M} and  \bibinfo{author}{{Johansen} A} (\bibinfo{year}{2014}), \bibinfo{month}{Dec.}
\bibinfo{title}{{Forming the cores of giant planets from the radial pebble flux in protoplanetary discs}}.
\bibinfo{journal}{{\em \aap}} \bibinfo{volume}{572}, \bibinfo{eid}{A107}. \bibinfo{doi}{\doi{10.1051/0004-6361/201424343}}.
\eprint{1408.6094}.

\bibtype{Article}%
\bibitem[{Malamud} et al.(2024)]{2024arXiv240312512M}
\bibinfo{author}{{Malamud} U}, \bibinfo{author}{{Podolak} M}, \bibinfo{author}{{Podolak} JI} and  \bibinfo{author}{{Bodenheimer} PH} (\bibinfo{year}{2024}), \bibinfo{month}{Oct.}
\bibinfo{title}{{Uranus and Neptune as methane planets: Producing icy giants from refractory planetesimals}}.
\bibinfo{journal}{{\em \icarus}} \bibinfo{volume}{421}, \bibinfo{eid}{116217}. \bibinfo{doi}{\doi{10.1016/j.icarus.2024.116217}}.
\eprint{2403.12512}.

\bibtype{Article}%
\bibitem[{Marley} et al.(1995)]{Marley1995}
\bibinfo{author}{{Marley} MS}, \bibinfo{author}{{G{\'o}mez} P} and  \bibinfo{author}{{Podolak} M} (\bibinfo{year}{1995}), \bibinfo{month}{Nov.}
\bibinfo{title}{{Monte Carlo interior models for Uranus and Neptune}}.
\bibinfo{journal}{{\em \jgr}} \bibinfo{volume}{100} (\bibinfo{number}{E11}): \bibinfo{pages}{23349--23354}. \bibinfo{doi}{\doi{10.1029/95JE02362}}.

\bibtype{Article}%
\bibitem[{Masters}(2018)]{2018GeoRL..45.7320M}
\bibinfo{author}{{Masters} A} (\bibinfo{year}{2018}), \bibinfo{month}{Aug.}
\bibinfo{title}{{A More Viscous-Like Solar Wind Interaction With All the Giant Planets}}.
\bibinfo{journal}{{\em \grl}} \bibinfo{volume}{45} (\bibinfo{number}{15}): \bibinfo{pages}{7320--7329}. \bibinfo{doi}{\doi{10.1029/2018GL078416}}.

\bibtype{Article}%
\bibitem[{Morf} et al.(2024)]{Morf2024}
\bibinfo{author}{{Morf} L}, \bibinfo{author}{{M{\"u}ller} S} and  \bibinfo{author}{{Helled} R} (\bibinfo{year}{2024}), \bibinfo{month}{Oct.}
\bibinfo{title}{{The interior of Uranus: Thermal profile, bulk composition, and the distribution of rock, water, and hydrogen and helium}}.
\bibinfo{journal}{{\em \aap}} \bibinfo{volume}{690}, \bibinfo{eid}{A105}. \bibinfo{doi}{\doi{10.1051/0004-6361/202450698}}.
\eprint{2408.10336}.

\bibtype{Article}%
\bibitem[Movshovitz and Fortney(2022)]{Movshovitz22}
\bibinfo{author}{Movshovitz N} and  \bibinfo{author}{Fortney J} (\bibinfo{year}{2022}).
\bibinfo{title}{{The Promise and Limitations of Precision Gravity: Application to the Interior Structure of Uranus and Neptune}}.
\bibinfo{journal}{{\em PSJ}} \bibinfo{volume}{3}: \bibinfo{pages}{88}.

\bibtype{Book}%
\bibitem[{National Research Council}(2022)]{NRC2022}
\bibinfo{author}{{National Research Council}} (\bibinfo{year}{2022}).
\bibinfo{title}{Origins, Worlds, and Life: A Decadal Strategy for Planetary Science and Astrobiology 2023-2032}, \bibinfo{publisher}{The National Academies Press}.
\bibinfo{doi}{\doi{10.17226/26522}}.

\bibtype{Article}%
\bibitem[{Nettelmann} et al.(2013)]{nettel13}
\bibinfo{author}{{Nettelmann} N}, \bibinfo{author}{{Helled} R}, \bibinfo{author}{{Fortney} JJ} and  \bibinfo{author}{{Redmer} R} (\bibinfo{year}{2013}), \bibinfo{month}{Mar.}
\bibinfo{title}{{New indication for a dichotomy in the interior structure of Uranus and Neptune from the application of modified shape and rotation data}}.
\bibinfo{journal}{{\em \planss}} \bibinfo{volume}{77}: \bibinfo{pages}{143--151}. \bibinfo{doi}{\doi{10.1016/j.pss.2012.06.019}}.
\eprint{1207.2309}.

\bibtype{Article}%
\bibitem[{Neuenschwander} and {Helled}(2022)]{Neuenschwander2022}
\bibinfo{author}{{Neuenschwander} BA} and  \bibinfo{author}{{Helled} R} (\bibinfo{year}{2022}), \bibinfo{month}{May}.
\bibinfo{title}{{Empirical structure models of Uranus and Neptune}}.
\bibinfo{journal}{{\em \mnras}} \bibinfo{volume}{512} (\bibinfo{number}{3}): \bibinfo{pages}{3124--3136}. \bibinfo{doi}{\doi{10.1093/mnras/stac628}}.
\eprint{2203.02233}.

\bibtype{Article}%
\bibitem[{Neuenschwander} et al.(2024)]{Neuenschwander2024}
\bibinfo{author}{{Neuenschwander} BA}, \bibinfo{author}{{M{\"u}ller} S} and  \bibinfo{author}{{Helled} R} (\bibinfo{year}{2024}), \bibinfo{month}{Jan.}
\bibinfo{title}{{Uranus' Complex Internal Structure}}.
\bibinfo{journal}{{\em arXiv e-prints}} , \bibinfo{eid}{arXiv:2401.11769}\bibinfo{doi}{\doi{10.48550/arXiv.2401.11769}}.
\eprint{2401.11769}.

\bibtype{Article}%
\bibitem[{Podolak} and {Helled}(2012)]{2012ApJ...759L..32P}
\bibinfo{author}{{Podolak} M} and  \bibinfo{author}{{Helled} R} (\bibinfo{year}{2012}), \bibinfo{month}{Nov.}
\bibinfo{title}{{What Do We Really Know about Uranus and Neptune?}}
\bibinfo{journal}{{\em \apjl}} \bibinfo{volume}{759} (\bibinfo{number}{2}), \bibinfo{eid}{L32}. \bibinfo{doi}{\doi{10.1088/2041-8205/759/2/L32}}.
\eprint{1208.5551}.

\bibtype{Article}%
\bibitem[{Podolak} et al.(1995)]{Podolak1995}
\bibinfo{author}{{Podolak} M}, \bibinfo{author}{{Weizman} A} and  \bibinfo{author}{{Marley} M} (\bibinfo{year}{1995}), \bibinfo{month}{Dec.}
\bibinfo{title}{{Comparative models of Uranus and Neptune}}.
\bibinfo{journal}{{\em \planss}} \bibinfo{volume}{43} (\bibinfo{number}{12}): \bibinfo{pages}{1517--1522}. \bibinfo{doi}{\doi{10.1016/0032-0633(95)00061-5}}.

\bibtype{Article}%
\bibitem[{Pollack} et al.(1996)]{Pollack1996}
\bibinfo{author}{{Pollack} JB}, \bibinfo{author}{{Hubickyj} O}, \bibinfo{author}{{Bodenheimer} P}, \bibinfo{author}{{Lissauer} JJ}, \bibinfo{author}{{Podolak} M} and  \bibinfo{author}{{Greenzweig} Y} (\bibinfo{year}{1996}), \bibinfo{month}{Nov.}
\bibinfo{title}{{Formation of the Giant Planets by Concurrent Accretion of Solids and Gas}}.
\bibinfo{journal}{{\em \icarus}} \bibinfo{volume}{124} (\bibinfo{number}{1}): \bibinfo{pages}{62--85}. \bibinfo{doi}{\doi{10.1006/icar.1996.0190}}.

\bibtype{Article}%
\bibitem[{Redmer} et al.(2011)]{redmer}
\bibinfo{author}{{Redmer} R}, \bibinfo{author}{{Mattsson} TR}, \bibinfo{author}{{Nettelmann} N} and  \bibinfo{author}{{French} M} (\bibinfo{year}{2011}), \bibinfo{month}{Jan.}
\bibinfo{title}{{The phase diagram of water and the magnetic fields of Uranus and Neptune}}.
\bibinfo{journal}{{\em \icarus}} \bibinfo{volume}{211} (\bibinfo{number}{1}): \bibinfo{pages}{798--803}. \bibinfo{doi}{\doi{10.1016/j.icarus.2010.08.008}}.

\bibtype{Article}%
\bibitem[{Reinhardt} et al.(2020)]{Reinhardt2020}
\bibinfo{author}{{Reinhardt} C}, \bibinfo{author}{{Chau} A}, \bibinfo{author}{{Stadel} J} and  \bibinfo{author}{{Helled} R} (\bibinfo{year}{2020}), \bibinfo{month}{Mar.}
\bibinfo{title}{{Bifurcation in the history of Uranus and Neptune: the role of giant impacts}}.
\bibinfo{journal}{{\em \mnras}} \bibinfo{volume}{492} (\bibinfo{number}{4}): \bibinfo{pages}{5336--5353}. \bibinfo{doi}{\doi{10.1093/mnras/stz3271}}.
\eprint{1907.09809}.

\bibtype{Article}%
\bibitem[{Reynolds} and {Summers}(1965)]{1965JGR....70..199R}
\bibinfo{author}{{Reynolds} RT} and  \bibinfo{author}{{Summers} AL} (\bibinfo{year}{1965}), \bibinfo{month}{Jan.}
\bibinfo{title}{{Models of Uranus and Neptune}}.
\bibinfo{journal}{{\em \jgr}} \bibinfo{volume}{70} (\bibinfo{number}{1}): \bibinfo{pages}{199--208}. \bibinfo{doi}{\doi{10.1029/JZ070i001p00199}}.

\bibtype{Article}%
\bibitem[{Ribas} et al.(2015)]{2015A&A...576A..52R}
\bibinfo{author}{{Ribas} {\'A}}, \bibinfo{author}{{Bouy} H} and  \bibinfo{author}{{Mer{\'\i}n} B} (\bibinfo{year}{2015}), \bibinfo{month}{Apr.}
\bibinfo{title}{{Protoplanetary disk lifetimes vs. stellar mass and possible implications for giant planet populations}}.
\bibinfo{journal}{{\em \aap}} \bibinfo{volume}{576}, \bibinfo{eid}{A52}. \bibinfo{doi}{\doi{10.1051/0004-6361/201424846}}.
\eprint{1502.00631}.

\bibtype{Article}%
\bibitem[{Saillenfest} et al.(2022)]{2022A&A...668A.108S}
\bibinfo{author}{{Saillenfest} M}, \bibinfo{author}{{Rogoszinski} Z}, \bibinfo{author}{{Lari} G}, \bibinfo{author}{{Bailli{\'e}} K}, \bibinfo{author}{{Bou{\'e}} G}, \bibinfo{author}{{Crida} A} and  \bibinfo{author}{{Lainey} V} (\bibinfo{year}{2022}), \bibinfo{month}{Dec.}
\bibinfo{title}{{Tilting Uranus via the migration of an ancient satellite}}.
\bibinfo{journal}{{\em \aap}} \bibinfo{volume}{668}, \bibinfo{eid}{A108}. \bibinfo{doi}{\doi{10.1051/0004-6361/202243953}}.
\eprint{2209.10590}.

\bibtype{Article}%
\bibitem[{Scheibe} et al.(2021)]{Scheibe2021}
\bibinfo{author}{{Scheibe} L}, \bibinfo{author}{{Nettelmann} N} and  \bibinfo{author}{{Redmer} R} (\bibinfo{year}{2021}), \bibinfo{month}{Jun.}
\bibinfo{title}{{Thermal evolution of Uranus and Neptune. II. Deep thermal boundary layer}}.
\bibinfo{journal}{{\em \aap}} \bibinfo{volume}{650}, \bibinfo{eid}{A200}. \bibinfo{doi}{\doi{10.1051/0004-6361/202140663}}.
\eprint{2105.01359}.

\bibtype{Article}%
\bibitem[Soderlund and Stanley(2020)]{krista2020}
\bibinfo{author}{Soderlund KM} and  \bibinfo{author}{Stanley S} (\bibinfo{year}{2020}).
\bibinfo{title}{The underexplored frontier of ice giant dynamos}.
\bibinfo{journal}{{\em Earth and Space Science Open Archive}} : \bibinfo{pages}{17}\bibinfo{doi}{\doi{10.1002/essoar.10503671.1}}.

\bibtype{Article}%
\bibitem[{Soyuer} et al.(2020)]{soyuer2020}
\bibinfo{author}{{Soyuer} D}, \bibinfo{author}{{Soubiran} F} and  \bibinfo{author}{{Helled} R} (\bibinfo{year}{2020}), \bibinfo{month}{Aug.}
\bibinfo{title}{{Constraining the depth of the winds on Uranus and Neptune via Ohmic dissipation}}.
\bibinfo{journal}{{\em MNRAS}} \bibinfo{volume}{498} (\bibinfo{number}{1}): \bibinfo{pages}{621--638}. \bibinfo{doi}{\doi{10.1093/mnras/staa2461}}.
\eprint{2008.05291}.

\bibtype{Article}%
\bibitem[{Soyuer} et al.(2023)]{soyuer2023}
\bibinfo{author}{{Soyuer} D}, \bibinfo{author}{{Neuenschwander} B} and  \bibinfo{author}{{Helled} R} (\bibinfo{year}{2023}), \bibinfo{month}{Jan.}
\bibinfo{title}{{Zonal Winds of Uranus and Neptune: Gravitational Harmonics, Dynamic Self-gravity, Shape, and Rotation}}.
\bibinfo{journal}{{\em Astronomical Journal}} \bibinfo{volume}{165} (\bibinfo{number}{1}), \bibinfo{eid}{27}. \bibinfo{doi}{\doi{10.3847/1538-3881/aca08d}}.
\eprint{2210.17389}.

\bibtype{Article}%
\bibitem[Sromovsky and Fry(2005)]{sromovsky}
\bibinfo{author}{Sromovsky L} and  \bibinfo{author}{Fry P} (\bibinfo{year}{2005}).
\bibinfo{title}{Dynamics of cloud features on uranus}.
\bibinfo{journal}{{\em Icarus}} \bibinfo{volume}{179} (\bibinfo{number}{2}): \bibinfo{pages}{459 -- 484}.
ISSN \bibinfo{issn}{0019-1035}. \bibinfo{doi}{\doi{https://doi.org/10.1016/j.icarus.2005.07.022}}.
\bibinfo{url}{\url{http://www.sciencedirect.com/science/article/pii/S0019103505002642}}.

\bibtype{Article}%
\bibitem[{Stixrude} et al.(2021)]{2021PSJ.....2..222S}
\bibinfo{author}{{Stixrude} L}, \bibinfo{author}{{Baroni} S} and  \bibinfo{author}{{Grasselli} F} (\bibinfo{year}{2021}), \bibinfo{month}{Dec.}
\bibinfo{title}{{Thermal and Tidal Evolution of Uranus with a Growing Frozen Core}}.
\bibinfo{journal}{{\em \psj}} \bibinfo{volume}{2} (\bibinfo{number}{6}), \bibinfo{eid}{222}. \bibinfo{doi}{\doi{10.3847/PSJ/ac2a47}}.

\bibtype{Article}%
\bibitem[{Teanby} et al.(2020)]{Teanby2020}
\bibinfo{author}{{Teanby} NA}, \bibinfo{author}{{Irwin} PGJ}, \bibinfo{author}{{Moses} JI} and  \bibinfo{author}{{Helled} R} (\bibinfo{year}{2020}), \bibinfo{month}{Dec.}
\bibinfo{title}{{Neptune and Uranus: ice or rock giants?}}
\bibinfo{journal}{{\em Philosophical Transactions of the Royal Society of London Series A}} \bibinfo{volume}{378} (\bibinfo{number}{2187}), \bibinfo{eid}{20190489}. \bibinfo{doi}{\doi{10.1098/rsta.2019.0489}}.

\bibtype{Inproceedings}%
\bibitem[Tian(2015)]{Tian2015PlanetaryDM}
\bibinfo{author}{Tian Y} (\bibinfo{year}{2015}), \bibinfo{title}{Planetary dynamos: Magnetic constraints on the interior structure and evolution of a planet}, \bibinfo{url}{\url{https://api.semanticscholar.org/CorpusID:123751568}}.

\bibtype{Article}%
\bibitem[{Tsiganis} et al.(2005)]{Tsiganis2005}
\bibinfo{author}{{Tsiganis} K}, \bibinfo{author}{{Gomes} R}, \bibinfo{author}{{Morbidelli} A} and  \bibinfo{author}{{Levison} HF} (\bibinfo{year}{2005}), \bibinfo{month}{May}.
\bibinfo{title}{{Origin of the orbital architecture of the giant planets of the Solar System}}.
\bibinfo{journal}{{\em \nat}} \bibinfo{volume}{435} (\bibinfo{number}{7041}): \bibinfo{pages}{459--461}. \bibinfo{doi}{\doi{10.1038/nature03539}}.

\bibtype{Article}%
\bibitem[{Valletta} and {Helled}(2022)]{Valletta2022}
\bibinfo{author}{{Valletta} C} and  \bibinfo{author}{{Helled} R} (\bibinfo{year}{2022}), \bibinfo{month}{May}.
\bibinfo{title}{{Possible In Situ Formation of Uranus and Neptune via Pebble Accretion}}.
\bibinfo{journal}{{\em \apj}} \bibinfo{volume}{931} (\bibinfo{number}{1}), \bibinfo{eid}{21}. \bibinfo{doi}{\doi{10.3847/1538-4357/ac5f52}}.
\eprint{2203.06545}.

\bibtype{Article}%
\bibitem[{Vazan} and {Helled}(2020)]{Vazan2020}
\bibinfo{author}{{Vazan} A} and  \bibinfo{author}{{Helled} R} (\bibinfo{year}{2020}), \bibinfo{month}{Jan.}
\bibinfo{title}{{Explaining the low luminosity of Uranus: a self-consistent thermal and structural evolution}}.
\bibinfo{journal}{{\em \aap}} \bibinfo{volume}{633}, \bibinfo{eid}{A50}. \bibinfo{doi}{\doi{10.1051/0004-6361/201936588}}.
\eprint{1908.10682}.

\bibtype{Article}%
\bibitem[{Zlimen} et al.(2024)]{2024AJ....168...64Z}
\bibinfo{author}{{Zlimen} E}, \bibinfo{author}{{Bailey} E} and  \bibinfo{author}{{Murray-Clay} R} (\bibinfo{year}{2024}), \bibinfo{month}{Aug.}
\bibinfo{title}{{Extensive Pollution of Uranus and Neptune's Atmospheres by Upsweep of Icy Material during the Nice Model Migration}}.
\bibinfo{journal}{{\em \aj}} \bibinfo{volume}{168} (\bibinfo{number}{2}), \bibinfo{eid}{64}. \bibinfo{doi}{\doi{10.3847/1538-3881/ad4c6a}}.
\eprint{2405.09621}.

\end{thebibliography*}

\end{document}